\documentclass[final,1p,times]{elsarticle} 
\usepackage{graphicx} 
\usepackage{amssymb} 
\usepackage{amsthm} 
\usepackage{lineno} 

\journal{Nuclear Physics A} 
\begin{document} 

\begin{frontmatter} 


\title{Eccentricity Fluctuation in Initial Conditions of Hydrodynamics}

\author{Tetsufumi Hirano$^{a}$ and Yasushi Nara$^{b}$}

\address[a]{Department of Physics, The University of Tokyo, Tokyo 113-0033, Japan}

\address[b]{Akita International University, Yuwa, Akita-city, 010-1292, Japan}

\begin{abstract} 
We study effects of eccentricity fluctuations
on the elliptic flow coefficient $v_2$ at mid-rapidity
in both Au+Au and Cu+Cu collisions at $\sqrt{s_{NN}}=200$\,GeV
by using a hybrid model that combines ideal hydrodynamics for
space-time evolution of the quark gluon plasma phase and a
hadronic transport model for the hadronic matter.
We find that the effect of eccentricity fluctuation is modest in semicentral
Au+Au collisions but significantly enhances $v_2$ in Cu+Cu collisions.
\end{abstract} 

\end{frontmatter} 



\section{Introduction}
One of the major discoveries at 
Relativistic Heavy Ion Collider (RHIC)
in Brookhaven National Laboratory
is that the elliptic flow \cite{Ollitrault}
is found to be as large as an ideal hydrodynamic prediction
for the first time
in relativistic heavy ion collisions \cite{reviews}.
Systematic studies revealed, however, that the reasonable agreement
between results from
ideal hydrodynamics and elliptic flow data
has been achieved only by a particular combination of dynamical modeling
\cite{HG05}.
If one looks at the comparison carefully,
one finds the agreement between hydrodynamic results and
the data
is not perfect. This would be due to an absence of
initial fluctuation effects \cite{MS03}.
Therefore, further investigation is indispensable toward better understanding
of the elliptic flow data and the quark gluon plasma (QGP).

\section{Initial Conditions}
For initial conditions in this study, we employ 
the Monte-Carlo version of both the Glauber model \cite{Miller:2007ri} and
the factorized Kharzeev-Levin-Nardi (fKLN) model \cite{Drescher:2006ca}
to generate the initial distribution of
entropy density
on an event-by-event basis.
We first calculate a transverse entropy density profile
$s_0(\mathbf{x}_{\perp}) = s(\tau = \tau_0, x, y, \eta_s = 0)$
in each sample
at an impact parameter $b$ for a given centrality,
where $\tau_0=0.6$ fm/$c$ is the initial time for the hydrodynamical
simulations.
Then  the variances of the profile are obtained from
\begin{eqnarray}
\sigma_x^2 & = & \langle x^2 \rangle - \langle x \rangle^2,\\
\sigma_y^2 & = & \langle y^2 \rangle - \langle y \rangle^2,\\
\sigma_{xy} & = & \langle xy \rangle - \langle x \rangle\langle y \rangle.
\end{eqnarray}
Here $\langle \cdots \rangle$ describes the average
over transverse plane by weighting the entropy density in a single sample:
\begin{equation}
\langle \cdots \rangle = \frac{\int d^2 x_\perp \cdots s_0(\mathbf{x}_\perp)}{\int d^2 x_\perp s_0(\mathbf{x}_\perp)}.
\end{equation}
The eccentricity with respect to the reaction plane,
the participant eccentricity,
and the corresponding transverse areas
can be defined \cite{phobosfluc}, respectively, as
\begin{eqnarray}
\varepsilon_{\mathrm{RP}} & = & 
  \frac{\sigma_y^2- \sigma_x^2}{\sigma_y^2+ \sigma_x^2},\quad
\varepsilon_{\mathrm{part}} \enskip = \enskip \frac{\sqrt{(\sigma_y^2- \sigma_x^2)^2+4\sigma_{xy}^2}}{\sigma_y^2+ \sigma_x^2},\\
S_{\mathrm{RP}} & = & \pi \sqrt{\sigma_x^2 \sigma_y^2}, \quad
S_{\mathrm{part}} \enskip = \enskip \pi \sqrt{\sigma_x^2 \sigma_y^2-\sigma_{xy}^2}. 
\label{eq:area}
\end{eqnarray}
The impact parameter vector and the true reaction plane
are not known experimentally.
So one can set an apparent frame of created matter shifted 
by $(x, y) = (\langle x \rangle, \langle y \rangle)$
and tilted by $\Psi$ from
the true frame in the transverse plane \cite{phobosfluc}:
\begin{eqnarray}
\tan 2 \Psi = \frac{\sigma_y^2-\sigma_x^2}{2\sigma_{xy}}.
\end{eqnarray}
The anisotropy of particle distribution
could be correlated with the misidentified frame.
To account for this,
we first shift the center-of-mass of the system
to the origin in the calculation frame and then
rotate the profile in the azimuthal direction by $\Psi$
to match the apparent reaction plane to
the true reaction plane.
We generate the next sample of an entropy density profile again as above
and average the profiles over many samples.
We repeat the above procedure for many samples
until the initial distribution is smooth enough.
The initial conditions obtained in this way 
contain the effects of eccentricity fluctuation
even though the profile is smooth.
In particular, even in case of vanishing impact parameter,
the eccentricity is finite due to its event-by-event fluctuation.
It is the particle distribution calculated from the initial conditions
mentioned above that can be directly compared with the experimental data. 
Note that the procedure of
averaging over many samples without shift or rotation
corresponds to conventional initialization
without the effect of
eccentricity fluctuation.

By using the initial conditions above, we simulate
space-time dynamics of matter produced in relativistic heavy ion collisions
with a hybrid approach in which
the macroscopic description of the QGP
is followed by the microscopic description of the hadron gas.
For details, see Ref.~\cite{HHKLN}.

\begin{figure}[ht]
\centering
\includegraphics[width=0.45\textwidth]{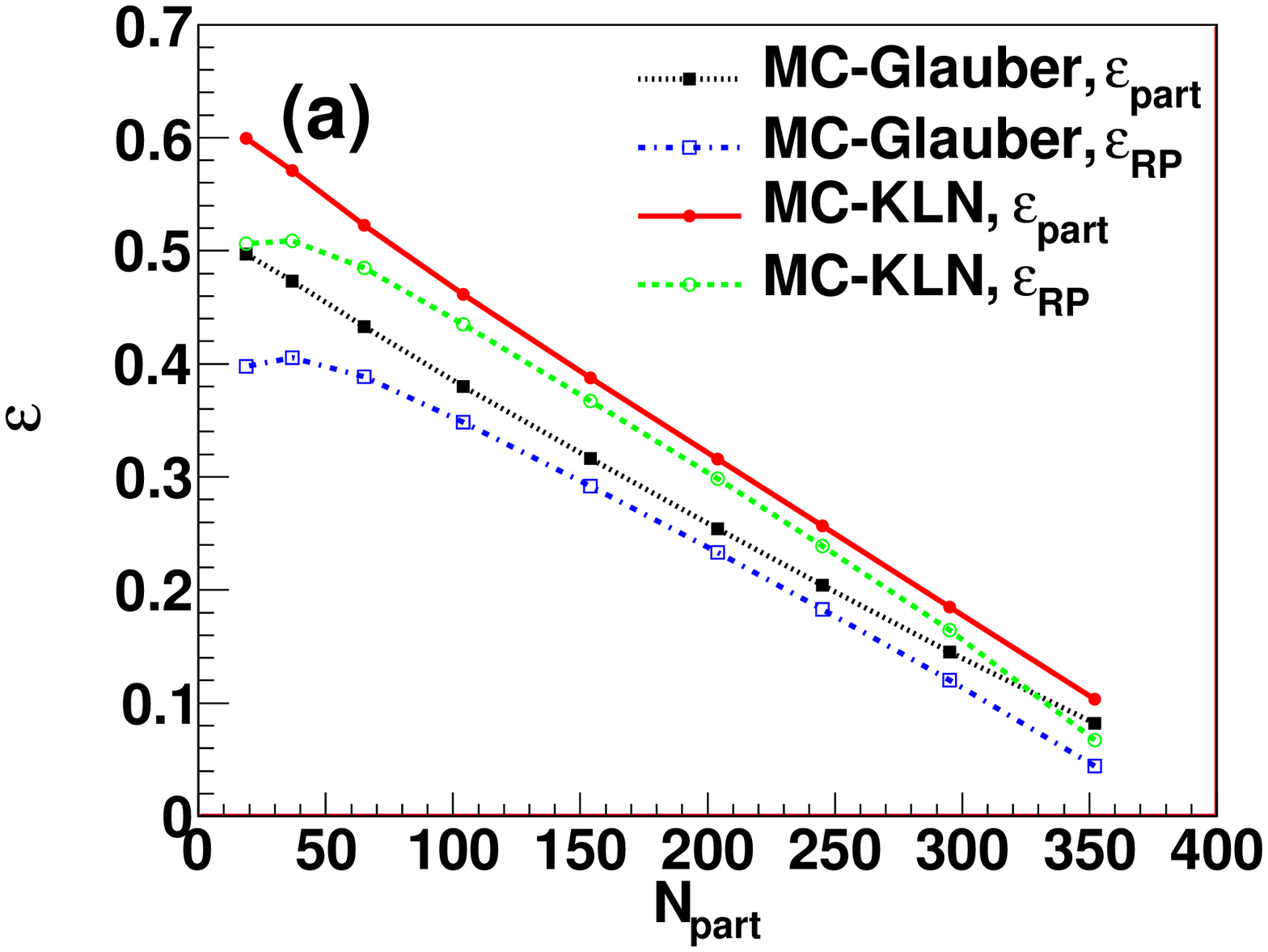}
\includegraphics[width=0.45\textwidth]{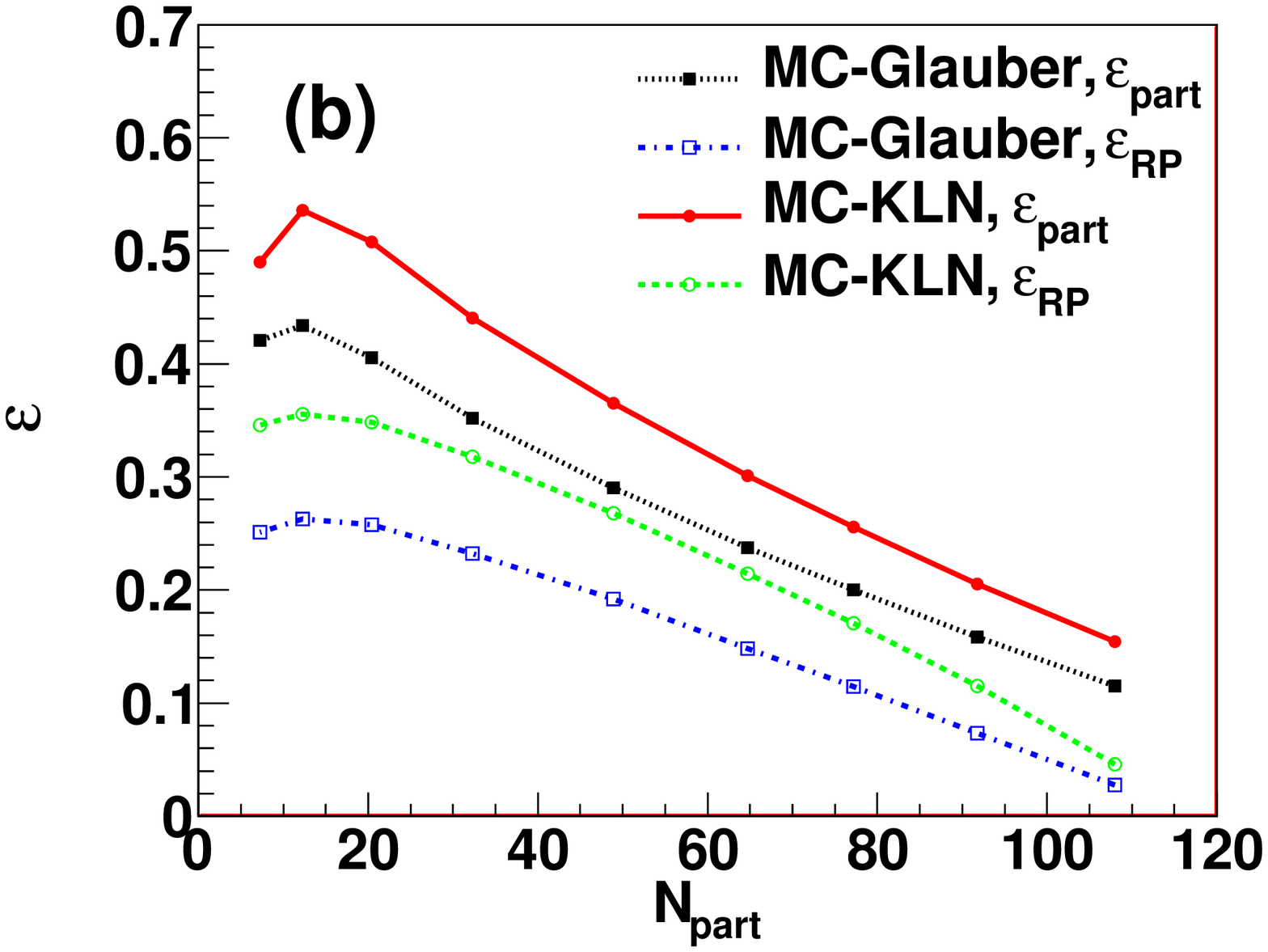}
\caption[]{
Eccentricity as a function of $N_{\mathrm{part}}$
in (a) Au+Au and (b) Cu+Cu collisions.
Solid, dotted, dashed and dash-dotted
lines correspond to
$\varepsilon_{\mathrm{part}}$ from the MC-KLN model,
$\varepsilon_{\mathrm{part}}$ from the MC-Glauber model,
$\varepsilon_{\mathrm{RP}}$ from the MC-KLN model
and
$\varepsilon_{\mathrm{RP}}$ from the MC-Glauber model,
respectively.
}
\label{fig:ecc}
\end{figure}

\section{Results}
Figure \ref{fig:ecc} shows 
the eccentricities as functions of 
the number of participants $N_{\mathrm{part}}$
with or without eccentricity fluctuations.
The results from the MC-Glauber and MC-KLN models
are compared with each other
in (a) Au+Au and (b) Cu+Cu collisions.
In semicentral Au+Au collisions at 10-50\% centrality,
the effect of initial fluctuations
enhances the eccentricity parameter by 8-11\% (5-8\%)
in the MC-Glauber (MC-KLN) model.
The enhancement factor
$\varepsilon_{\mathrm{part}}/\varepsilon_{\mathrm{RP}}$
is the largest at the very central bin (0-5\% centrality):
$\varepsilon_{\mathrm{part}}/\varepsilon_{\mathrm{RP}} =$ 1.83
in the MC-Glauber model and 1.53 in the MC-KLN model.
A qualitatively similar behavior is observed
in Cu+Cu collisions, but quantitatively the effect of fluctuation 
is significant due to the smallness of the system.

\begin{figure}[ht]
\centering
\includegraphics[width=0.45\textwidth]{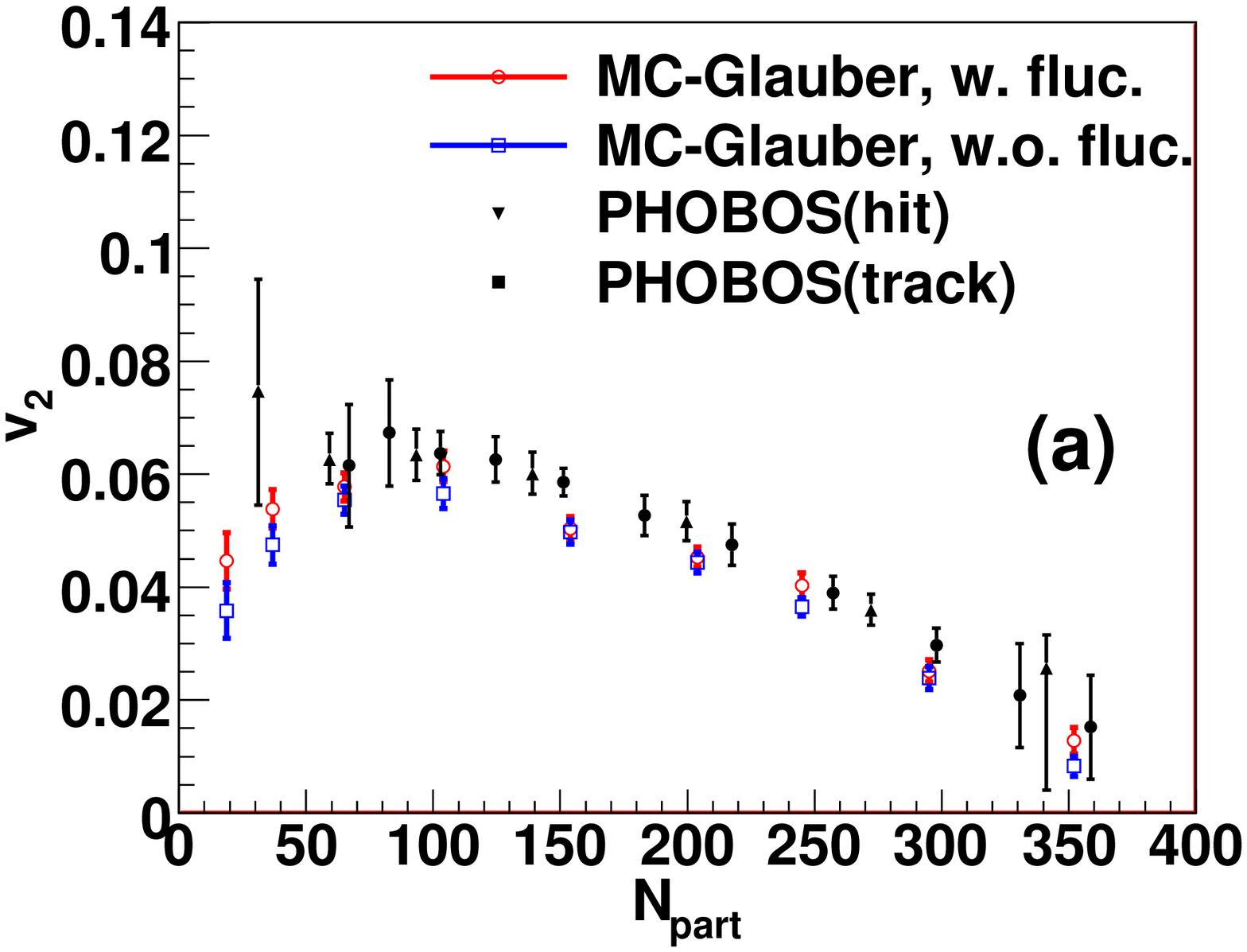}
\includegraphics[width=0.45\textwidth]{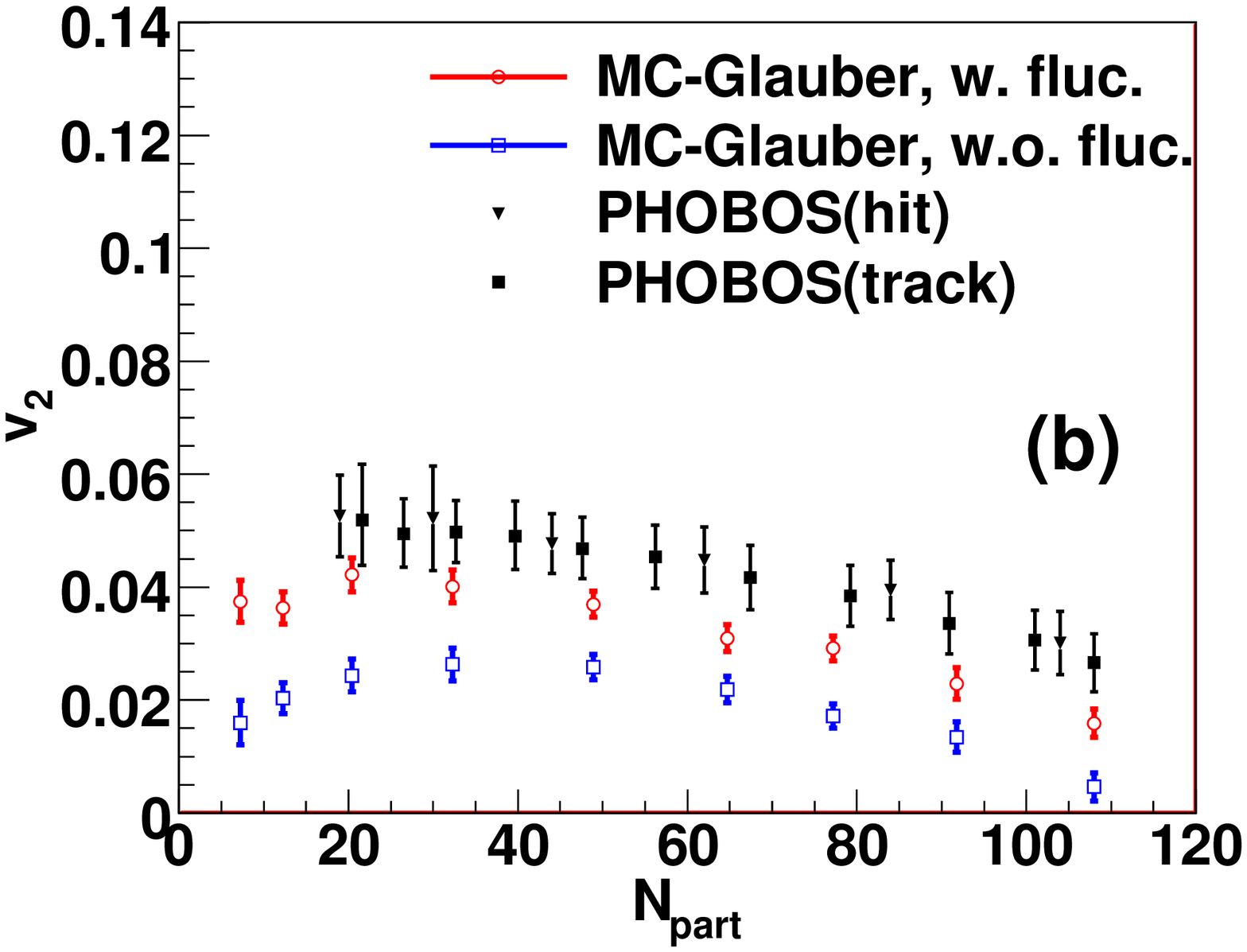}
\caption[]{
Centrality dependences of $v_2$ for charged particles
at mid-rapidity in (a) Au+Au and (b) Cu+Cu collisions
at $\sqrt{s_{NN}} = 200$ GeV with the MC-Glauber model initial conditions
are compared with PHOBOS data (filled plots) \cite{Back:2004mh}.
Open circles (squares) are results with (without) eccentricity fluctuation.
}
\label{fig:v2glauber}
\end{figure}
\begin{figure}[ht]
\centering
\includegraphics[width=0.45\textwidth]{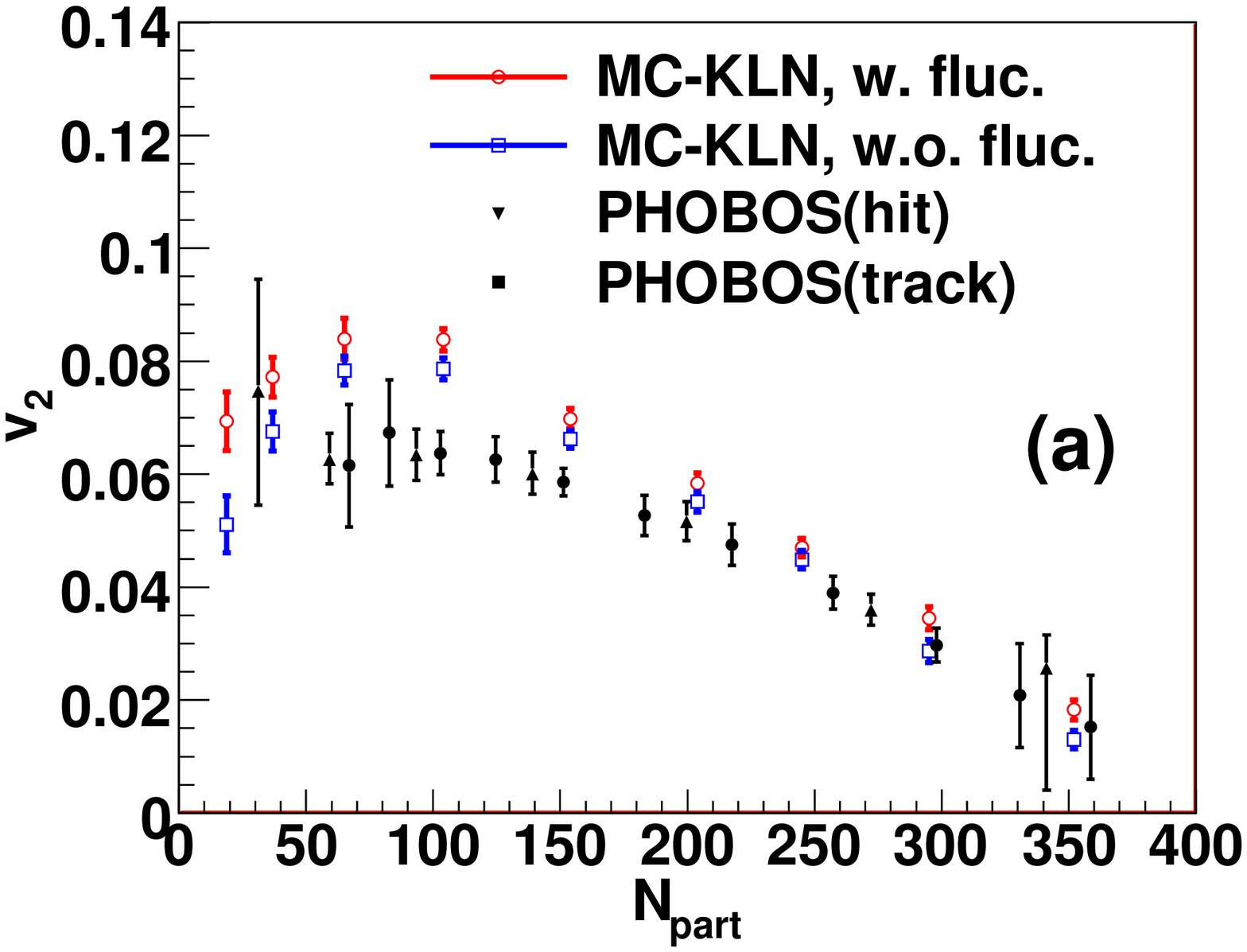}
\includegraphics[width=0.45\textwidth]{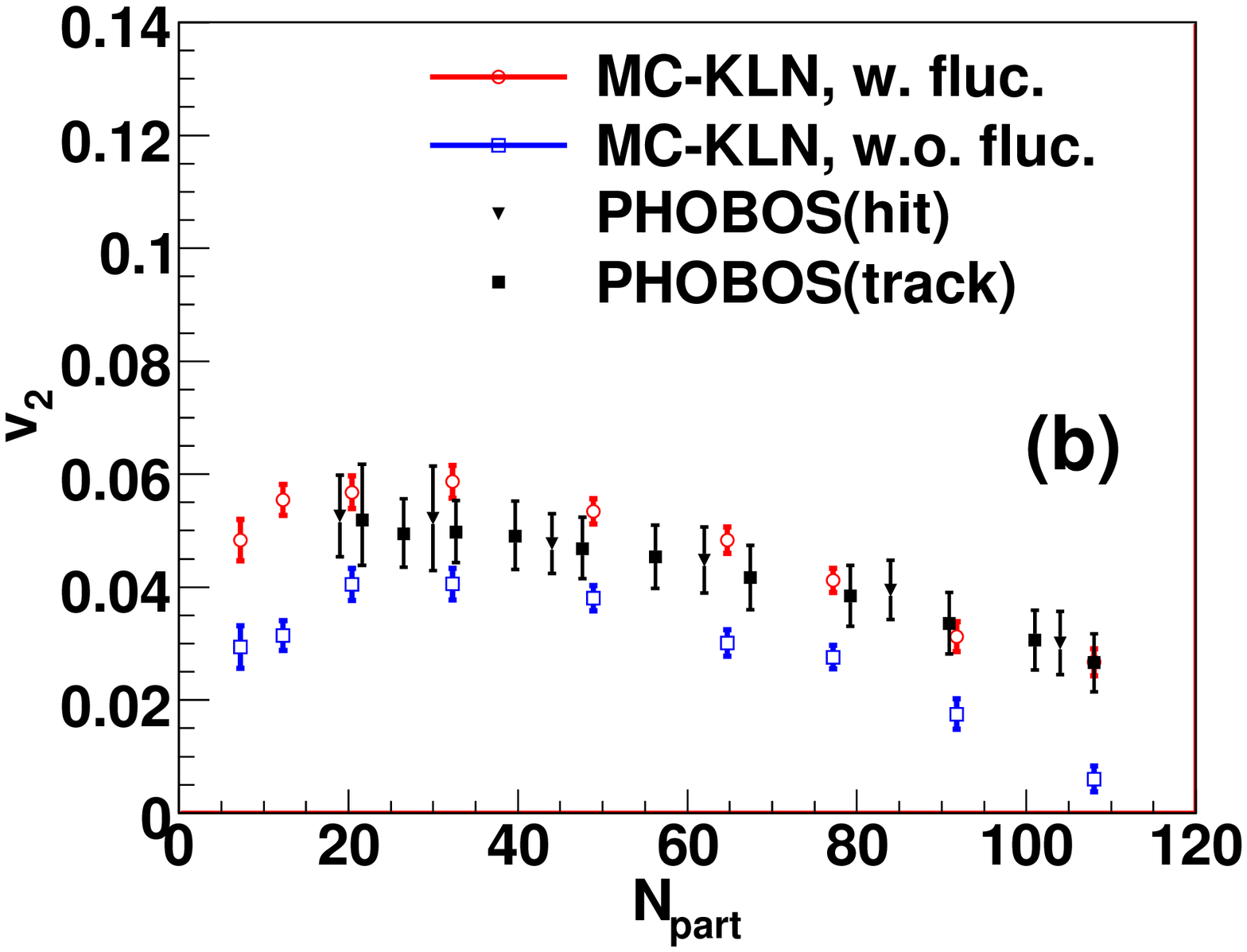}
\caption[]{
Centrality dependences of $v_2$ for charged particles
at mid-rapidity in (a) Au+Au and (b) Cu+Cu collisions
at $\sqrt{s_{NN}} = 200$ GeV with the MC-KLN model initial conditions
are compared with PHOBOS data (filled plots) \cite{Back:2004mh}.
Open circles (squares) are results with (without) eccentricity fluctuation.
}
\label{fig:v2kln}
\end{figure}

Figures \ref{fig:v2glauber}
and \ref{fig:v2kln}
show the centrality dependence of
$v_2$ for charged hadrons
at mid-rapidity ($\mid \eta \mid < 1$)
in Au+Au and Cu+Cu collisions at $\sqrt{s_{NN}} = 200$ GeV.
From Figs.~\ref{fig:v2glauber} (a) and (b),
one sees that the effect of eccentricity fluctuations
is not significant in Au+Au collisions,
whereas
$v_2$ is largely enhanced 
in Cu+Cu collisions due to fluctuation effects.
These tendencies are also expected from the results of
initial eccentricity as shown in Fig.~\ref{fig:ecc}.
It is interesting to point out that elliptic flow coefficients 
$v_2$ in the Glauber model initial conditions still slightly undershoot
the experimental data, particularly in Cu+Cu collisions,
 even with fluctuation effects.
Thus there is almost no room for the viscosity 
in the QGP stage to play a role
in the Glauber initial conditions within
the hybrid approach with the ideal gas equation of state in the QGP phase.
Figure \ref{fig:v2kln} is the same as Fig.~\ref{fig:v2glauber}
but the initial conditions are taken from the MC-KLN model.
Again, a similar qualitative behavior
is seen: The effect of eccentricity fluctuations is small
in Au+Au collisions but is large in Cu+Cu collisions.
Due to larger initial eccentricity in the MC-KLN model
than the MC-Glauber model,
the results are somewhat larger than the experimental data
in peripheral Au+Au collisions,
whereas we reasonably reproduce
the $v_2$ data with initial fluctuation effects
in Cu+Cu collisions.
Viscous effects in the QGP phase
could reduce the $v_2$ and
enable us to reproduce
the data in Au+Au collisions
in this case.
However, the results
are already comparable with
the data in Cu+Cu collisions
even though the number of participants is almost the same
as that in peripheral Au+Au collisions.
So it would be non-trivial to establish whether
the same viscous effects 
also give the right amount of $v_2$
in Cu+Cu collisions.

\section{Conclusions}
We calculated the elliptic flow
coefficient as a function of the number of participants
in the QGP hydro plus the hadronic cascade model
and found that the effect of eccentricity fluctuations is 
visible in very central and peripheral Au+Au collisions and 
is quite large in Cu+Cu collisions.
This strongly suggests that the effect of eccentricity fluctuations
is an important factor which has to be included in the dynamical model
for understanding of the elliptic flow data
and for precise extraction
of transport properties of the produced matter from the data.

So far, one has been focusing on comparison of hydrodynamic results
with $v_2$ data only in Au+Au collisions.
The experimental data in Cu+Cu collisions
also have a strong power to constrain the 
dynamical models. Therefore,
simultaneous analysis of 
$v_2$ data in both Au+Au and Cu+Cu collisions
will be called for in future
hydrodynamic studies.


\section*{Acknowledgments}
The work of T.H. was partly supported by Grant-in-Aid for Scientific Research No.~19740130 and by Sumitomo Foundation No.~080734. 
The work of Y.N. was supported by Grand-in-Aid for Scientific
Research No.~20540276.


\begin{thebibliography}{00} 

\bibitem{Ollitrault}
  J.~Y.~Ollitrault,
  Phys.\ Rev.\ D 46 (1992) 229.

\bibitem{reviews}
  T.~Hirano, N.~van der Kolk, A.~Bilandzic,
  arXiv:0808.2684 [nucl-th];

\bibitem{HG05}
  T.~Hirano, M.~Gyulassy,
  Nucl.\ Phys.\ A 769 (2006) 71.

\bibitem{MS03}
  M.~Miller, R.~Snellings,
  nucl-ex/0312008.

\bibitem{Miller:2007ri}
  M.~L.~Miller, K.~Reygers, S.~J.~Sanders, P.~Steinberg,
  Ann.\ Rev.\ Nucl.\ Part.\ Sci.\ 57 (2007) 205.

\bibitem{Drescher:2006ca}
  H.~J.~Drescher, Y.~Nara,
  Phys.\ Rev.\  C 75 (2007) 034905;

  H.~J.~Drescher, Y.~Nara,
  Phys.\ Rev.\  C 76 (2007) 041903.

\bibitem{phobosfluc}
  B.~Alver {\it et al.} [PHOBOS Collaboration], 
  nucl-ex/0702036; 

 B.~Alver {\it et al.},
  Phys.\ Rev.\  C 77 (2008) 014906.

\bibitem{HHKLN}
  T.~Hirano, U.~Heinz, D.~Kharzeev, R.~Lacey, Y.~Nara,
  Phys.\ Lett.\ B 636 (2006) 299;

T.~Hirano, U.~Heinz, D.~Kharzeev, R.~Lacey, Y.~Nara,
  Phys.\ Rev.\  C 77 (2008) 044909.

\bibitem{Back:2004mh}
  B.~B.~Back {\it et al.}  [PHOBOS Collaboration],
  Phys.\ Rev.\  C 72 (2005) 051901;

  B.~Alver {\it et al.}  [PHOBOS Collaboration],
  Phys.\ Rev.\ Lett.\  98 (2007) 242302.

\end{thebibliography}
\end{document}